\documentclass[aps,amsmath,amssymb,amsfonts,showpace,twocolumn,pra]{revtex4}

\usepackage{amscd}
\usepackage{amsthm}
\usepackage{graphicx}
\usepackage{mathdots}

\usepackage[
  colorlinks,
  linkcolor = blue,
  citecolor = blue,
  urlcolor = blue]{hyperref}

\usepackage{amsmath,amssymb,amsthm}

\usepackage{tikz}
\usetikzlibrary{arrows}
\usetikzlibrary{calc,positioning}

\usepackage{thm-restate}

\usepackage{colonequals}

\usepackage[font=small]{caption}

\newtheoremstyle{noCaption}
{\topsep}
{\topsep}
{\itshape}
{}
{}
{}
{0pt}
{}%

\def \qed {\hfill \vrule height7pt width 7pt depth 0pt}

\begin{document}
\title{Nonlocality of orthogonal product basis quantum states}
\author{Yan-Ling Wang$^{1}$, Mao-Sheng Li$^{1}$,   Zhu-Jun Zheng$^{1}$ and Shao-Ming Fei$^{2,3}$}

 \affiliation
 {
 {\footnotesize  {$^1$Department of Mathematics,
 South China University of Technology, Guangzhou
510640, China}} \\
{\footnotesize{
  $^2$School of Mathematical Sciences, Capital Normal University,
Beijing 100048, China}}\\
{\footnotesize{$^3$Max-Planck-Institute for Mathematics in the Sciences, 04103
Leipzig, Germany}}
}

\begin{abstract}
We study the local indistinguishability of mutually orthogonal product basis quantum states in the high-dimensional quantum system. In the quantum system of $\mathbb{C}^d\otimes\mathbb{C}^d$, where $d$ is odd, Zhang \emph{et al} have constructed $d^2$ orthogonal product basis quantum states which are locally indistinguishable in [Phys. Rev. A. {\bf 90}, 022313(2014)]. We find a subset contains with $6d-9$ orthogonal product states which are still locally indistinguishable. Then we generalize our method to arbitrary bipartite quantum
system $\mathbb{C}^m\otimes\mathbb{C}^n$. We  present a small set with only $3(m+n)-9$ orthogonal product states and  prove these states are LOCC indistinguishable.  Even though these $3(m+n)-9$ product states
are LOCC indistinguishable, they can be distinguished by separable measurements. This shows that
separable operations are strictly stronger than the local operations and classical communication.
\end{abstract}

\maketitle

\section{Introduction}
Many global operators  can not be implemented using only  local operations and classical communication (LOCC).  Many researchers aim at studying  the limitation of quantum operators that can be implemented by LOCC. And the local distinguishability of quantum states plays an important role for the study of the limitation of LOCC \cite{Ben99}. Moreover, the  local indistinguishability of pure product states exhibits the phenomenon of nonlocality without entanglement\cite{Wal02,Ben99}.
Suppose Alice and Bob share a bipartite quantum system. They are not told which state their combined system is in, but they know it has been chosen from a specific set of mutually orthogonal states, that set being known to each of them. Their task is to identity  the given  state by using only LOCC. Distinguishing is essentially primitive for many quantum information tasks, such as quantum cryptography \cite{Ben92} and quantum algorithms \cite{Ber03}.

Recently, lots of authors paid attentions to the local distinguishablity of quantum states. Some of them considered the set with maximally entangled states \cite{Gho01,Fan04,Nathanson05,Fan07,Cohen07,Bandyopadhyay11,Yu12,Cosentino13,Yu14,Li15}, while the others aimed at the set with product states \cite{Wal02,Hor03,Rinaldis04,Che04,Gheorghiu08,Feng09,Duan09,Duan10,Yu11,Yang13,Zhang14,Ma14,Yu115}. Both of these researches can lead us a better understanding the limitation of the local operations and classical communication. Because the structure  of LOCC operations is mathematically complicated, some researchers investigated the difference limitation between the LOCC operations and the separable operations \cite{Chitambar09,Chitambar12,Childs13}. Clearly, every local operator is a separable operator. So it's natural to ask whether the separable operations are strictly stronger than local operations? In $\mathbb{C}^3\otimes\mathbb{C}^3$, the answer is positive and was shown in the paper of Bennett \emph{et al }\cite{Ben99}. In this paper, we give a positive answer to this question for high-dimensional system.

 In 1999, Bennett \emph{et al }\cite{Ben99} first presented  nine product states in $\mathbb{C}^3\otimes\mathbb{C}^3$ and proved that it cannot be distinguished by LOCC. This is the first example to exhibit the nonlocality property without entanglement. After that, Walgate \emph{et al} gave a sufficient and necessary condition for  the distinguishablity of orthogonal states in $\mathbb{C}^2\otimes\mathbb{C}^n$, and used it to prove the LOCC indistinguishability of the  nine product states in $\mathbb{C}^3\otimes\mathbb{C}^3$ \cite{Wal02}. Since then, the authors in \cite{Che04,Rinaldis04,Hor03,Duan09, Ma14} studied the LOCC distinguishability of pure product states. More recently, in \cite{Zhang14}, the authors generalized the result  to high-dimensional system and constructed $d^2$ orthogonal product basis quantum states which are locally indistinguishable in $\mathbb{C}^d\otimes\mathbb{C}^d$, where $d$ is odd.

The unextendible product bases (UPB) is helpful for us to generate an entangled state with partial positive transpose(PPT) \cite{Bennett99,DiVincenzo03,Alon01,Feng06,Niset06,Chen15}. It is well known that the set of
UPB constitutes a special class of locally indistinguishable product states \cite{Rinaldis04,Bennett99}. So it is interesting to ask whether there are some other classes of locally indistinguishable orthogonal product states except the UPB. This paper gives a positive answer to this question. Recently, the authors in \cite{Bandyopadhyay14} gave a necessary and sufficient condition for an UPB to be distinguished by separable measurements. So in this paper we also consider the distinguishablity of the set of product states by separable measurements.

 In this paper, we concentrate ourselves on finding the set  of  LOCC indistinguishable product states in $\mathbb{C}^m\otimes\mathbb{C}^n$. If a set of quantum states are LOCC distinguishable, then there should be a nontrivial measurement that can preserve the orthogonality of these states. If not, these quantum states must be LOCC indistinguishable. As in Ref.\cite{Zhang14}, the authors asked whether a subset of the quantum base they constructed is also LOCC indistinguishable. And we give an affirm answer to this question. First, we give an example. In $\mathbb{C}^7\otimes\mathbb{C}^7$,
the LOCC indistinguishable subset we present only contains 33 states. Then we generalize it to higher odd dimensions. These give a complete answer to the question asked by the authors in \cite{Zhang14}. Then we generalize our method to arbitrary bipartite quantum system and  give a small set of LOCC indistinguishable product states which contains only $3(m+n)-9$ states in $\mathbb{C}^m\otimes\mathbb{C}^n$. Then we prove the set of product states we constructed is not an UPB. Even though these product states are LOCC indistinguishable, they can be distinguished by separable measurements. This shows that separable operations are stronger than the local operations and classical communication.

\section{Small set of LOCC Indistinguishable Product States }
In this section, we first construct a subset (denote as $T$) of LOCC indistinguishable orthogonal product basis quantum states in $\mathbb{C}^d\otimes\mathbb{C}^d$, where d is odd. For a better understanding, we first show our construction in $\mathbb{C}^7\otimes\mathbb{C}^7$ as an example, then we generalize it to higher dimensions. In this paper, we use a rectangle to represent a quantum states. A rectangle contains two squares always presents two states: if the rectangle consists of the $i^{\text{th}}$ row, $j^{\text{th}}$ and $(j+1)^{\text{th}}$  columns, then the rectangle represents the two states $|i\rangle(|j\rangle\pm|j+1\rangle)$. And a rectangle contains only one square always presents one state: if the square consists of the $i^{\text{th}}$ row and  $j^{\text{th}}$ column, then the square represents the state $|i\rangle|j\rangle$. We make an agreement  that the set $T$ of product states are only chosen from the rectangle with grey color or black color. If the rectangle is
black, we choose both of the represented states, if the rectangle is grey we only choose one (the positive one), so the rectangle with only one square being chosen is always grey. For example, the first one rectangle  of the first row in \textbf{Fig.1} represents two states $|1\rangle(|1\rangle \pm|2\rangle)$, while the second one of the first
row represents the state $|1\rangle(|3\rangle+|4\rangle)$. And the square in the center of \textbf{Fig.1} represents the state $|4\rangle|4\rangle$. There
are 12 rectangles with black color and 9 rectangles with grey color, so the set $T$  contains $12\times2+9\times 1=33$ product states .

$$
\begin{tikzpicture}[
  node distance = 3pt,
  domino/.style = {rectangle, rounded corners = 2mm, draw = black!95, fill = black!20},
  dominoo/.style = {rectangle, rounded corners = 2mm, draw = black!95, fill = black!100},
  ]
  \def\step{20pt};   
  \def\dshort{16pt}; 
  \def\dlong{36pt};  
  \draw[step = \step, red, thin] (0,0) grid (7*\step, 7*\step);
 \foreach \i/\j in {6.5/6,5.5/5,4.5/4,2.5/3,1.5/2,0.5/1}
    \node[dominoo, double, minimum height = \dlong,  minimum width = \dshort] at (\i*\step, \j*\step) {};
 \foreach \i/\j in {6.5/4,5.5/3,1.5/4,0.5/3}
    \node[domino, double, minimum height = \dlong,  minimum width = \dshort] at (\i*\step, \j*\step) {};
  \node[domino, minimum height = \dshort,double, minimum width = \dshort] at (3.5*\step, 3.5*\step) {};
   \foreach \i/\j/\h in {1/6.5/3,2/5.5/3,3/4.5/3,4/2.5/3,5/1.5/3,6/0.5/3}
    \node[dominoo, double , minimum height = \dshort, minimum width = \dlong]  at (\i*\step, \j*\step) {};
   \foreach \i/\j/\h in {3/6.5/3,4/5.5/3,3/1.5/3,4/0.5/3}
    \node[domino, double , minimum height = \dshort, minimum width = \dlong]  at (\i*\step, \j*\step) {};
   \foreach \i in {1,2,3,4,5,6,7}{
    \node (a\i) at (\i*\step-0.5*\step, 5.3)  {$|\i\rangle$};
    \node (b\i) at (-0.35,-\i*\step+7.5*\step) {$|\i\rangle$};
  };
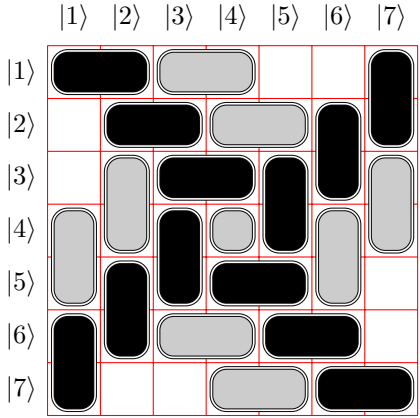
 \node [below = of {2.5,0}] {{\bf Fig.1} \text{Product states representation in } $\mathbb{C}^{7}\bigotimes\mathbb{C}^{7}$};
\end{tikzpicture}
$$

\bigskip
\noindent{\emph{Example}}:
 In $\mathbb{C}^{7}\bigotimes\mathbb{C}^{7},$ the $33$ states defined in the \textbf{Fig.1} are LOCC indistinguishable.

\emph{Proof:} Since these states are symmetrical, we know that if the states cannot be distinguished with Alice going first, then these states cannot be distinguished with Bob going first either. Thus, we only need to prove that these states cannot be distinguished by LOCC with Alice going first.

Suppose Alice do the first measurement $\{M_i^A\}$. Next
we prove that for any matrix $M$ that preserves the orthogonality of these states we constructed, then $M^\dagger M\propto \alpha I$ for
some $\alpha \in \mathbb{R}$.
Suppose $M^\dagger M$ is the following form under the basis $\{|1\rangle, |2\rangle,\ldots,|7\rangle \}.$
$$M^\dagger M=\left(
  \begin{array}{cccc}
    m_{11} & m_{12} & \cdots & m_{17} \\
    m_{21} & m_{22} & \cdots & m_{27} \\
    \vdots & \vdots & \ddots & \vdots \\
    m_{71} & m_{72} & \cdots & m_{77} \\
  \end{array}
\right).
$$
We have $7$ states
$$
\begin{array}{l}
|\psi_1\rangle=|1\rangle(|3\rangle+|4\rangle),
|\psi_2\rangle=|2\rangle(|4\rangle+|5\rangle),\\
|\psi_3\rangle=|3\rangle(|3\rangle+|4\rangle),
|\psi_4\rangle=|4\rangle|4\rangle,\\
|\psi_5\rangle=|5\rangle(|4\rangle+|5\rangle),
|\psi_6\rangle=|6\rangle(|3\rangle+|4\rangle),\\
|\psi_7\rangle=|7\rangle(|4\rangle+|5\rangle).
\end{array}
$$
If $\{|\psi_i\rangle\}_{i=1}^{33}$ are LOCC distinguishable, then $\{M \otimes I |\psi_j\rangle\}_{i=1}^{33}$ are pairwise orthogonal. Particularly, for the above $7$ states, we have
$$0=\langle\psi_i|M^\dagger M \otimes I |\psi_j\rangle=\langle i|M^\dagger M|j\rangle \langle \phi_i |\phi_j\rangle=m_{ij},\ \ i\neq j.$$
Because $|\phi_i\rangle=|3\rangle+|4\rangle,|4\rangle+|5\rangle, or |4\rangle$,
so $\langle \phi_i |\phi_j\rangle=1$, then the third equality of the above equation holds. So we have $m_{ij}=0$ for $i\neq j$.

Now we consider
$|\phi_1\rangle=(|1\rangle+|2\rangle)|7\rangle $ and
$|\phi_2\rangle=(|1\rangle-|2\rangle)|7\rangle .$
By the orthogonality of $(M\otimes I)|\phi_1\rangle \text{ and }(M\otimes I)|\phi_2\rangle$, we get
$$\langle \phi_1|M^{\dagger}M\otimes I|\phi_2\rangle=m_{11}+m_{21}-m_{12}-m_{22}=0.$$
Since $m_{ij}=0  \text{ for } \  i\neq j$, so we get $m_{11}=m_{22}.$ If we consider
$|\phi_3\rangle=(|2\rangle+|3\rangle)|6\rangle $ and
$|\phi_4\rangle=(|2\rangle-|3\rangle)|6\rangle $, we get $m_{22}=m_{33}$. Similarly, we can obtain $m_{33}=m_{44},m_{44}=m_{55},m_{55}=m_{66},m_{66}=m_{77}.$
Hence $M^\dagger M=diag(a,a,\ldots,a)=aI$ for some $a\in\mathbb{R}.$

In order to preserve the orthogonality of our given
states, any measurement $\{M_i^A\}$ Alice can do is the trivial
measurement. This implies the LOCC indistinguishability of the orthogonal product states we constructed.
\qed

$$
\begin{tikzpicture}[
  node distance = 2pt,
  domino/.style = {rectangle, rounded corners = 2mm, draw = black!95, fill = black!20},
  dominoo/.style = {rectangle, rounded corners = 2mm, draw = black!95, fill = black!100}]
  \def\step{15pt};   
  \def\dshort{12pt}; 
  \def\dlong{27pt};  
  \draw[step = \step, red, thin] (0,0) grid (13*\step, 13*\step);
   \foreach \i/\j in {12.5/12,11.5/11,9.5/9,8.5/8,7.5/7,5.5/6,4.5/5,3.5/4,1.5/2,0.5/1}
    \node[dominoo, double, minimum height = \dlong,  minimum width = \dshort] at (\i*\step, \j*\step) {};
       \foreach \i/\j in {0.5/6,1.5/7,3.5/6,4.5/7,8.5/6,9.5/7,11.5/6,12.5/7}
    \node[domino, double, minimum height = \dlong,  minimum width = \dshort] at (\i*\step, \j*\step) {};
   \node[domino, minimum height = \dshort,double, minimum width = \dshort] at (6.5*\step, 6.5*\step) {};
   \foreach \i/\j/\h in {1/12.5/3,2/11.5/3,4/9.5/3,5/8.5/3,6/7 .5/3,7/5.5/3,8/4.5/3,9/3.5/3,11/1.5/3,12/0.5/3}
    \node[dominoo, double , minimum height = \dshort, minimum width = \dlong]  at (\i*\step, \j*\step) {};
   \foreach \i/\j/\h in {6/12.5/3,7/11.5/3,6/9.5/3,7/8.5/3,6/4.5/3,7/3.5/3,6/1.5/3,7/0.5/3}
    \node[domino, double , minimum height = \dshort, minimum width = \dlong]  at (\i*\step, \j*\step) {};
  \node [below = of {3.5,0}] {{\bf Fig.2} \text{Product states representation in } $\mathbb{C}^{d}\bigotimes\mathbb{C}^{d}$};
  \node [below = of {1.5,6.1}] {$\ddots$};
  \node [below = of {3.4,6.1}] {$\vdots$};
  \node [below = of {1.35,3.6}] {$\cdots$};
  \node [below = of {3.4,1.9}] {$\vdots$};
  \node [below = of {5.6,3.6}] {$\cdots$};
  \node [below = of {5.4,1.9}] {$\ddots$};
  \node [below = of {5.5,5.9}] {$\iddots$};
   \node [below = of {1.35,2.1}] {$\iddots$};
\end{tikzpicture}
$$

\emph{Theorem 1.}
 In $\mathbb{C}^{d}\bigotimes\mathbb{C}^{d},$ where $d$ is odd. The $6d-9$ states defined in the \textbf{Fig.2} is LOCC indistinguishable. And we denote the set of product states we constructed as $T$ .

\emph{Proof:} We define the middle horizon states in   \textbf{Fig.2} from upper to below as $|\psi_i\rangle,\ \ i=1,2,\ldots,2n+1$ and the black vertical states
$$
\begin{array}{c}
|\phi_i^{\pm}\rangle=(|i\rangle\pm|i+1\rangle)|i\rangle ,i=1,2,\ldots,n,\\
|\phi_i^{\pm}\rangle=(|i-1\rangle\pm|i\rangle)|i\rangle,i=n+1,n+2,\ldots,2n+1.
\end{array}
$$

Suppose these states can be LOCC distinguished, then either Alice or Bob can perform a nontrivial measurement $\{M_i^A\}$ that can preserve the orthogonality of these states. As the states we constructed are symmetry, so we can suppose Alice do the first measurement $\{M_i^A\}$. We prove that for any matrix $M$ that preserves the orthogonality of $|\psi_i\rangle$and $|\phi_i^{\pm}\rangle$ , then $M^{\dagger}M\propto\alpha I$ for some $\alpha\in \mathbb{R}.$

If $M$ is one of such matrix, we can suppose $M^{\dagger}M=\{m_{ij}\}_{i,j=1}^d$. From the orthogonality of $M\otimes I|\psi_i\rangle$ and $M\otimes I|\psi_j\rangle$, we have
 $$0=\langle\psi_i|M^{\dagger}M|\psi_j\rangle=m_{ij}, \ \ \ i\neq j .$$
Now for the orthogonality between  $M\otimes I|\phi_i^{+}\rangle$ and $M\otimes I|\phi_i^{-}\rangle$, we have
{\small
$$
\begin{array}{l}
0=\langle\phi_i^{+}|M^{\dagger}M\otimes I|\phi_i^-\rangle=m_{ii}-m_{i,i+1}+m_{i+1,i}-m_{i+1,i+1},\\
0=\langle\phi_{j}^{+}|M^{\dagger}M\otimes I|\phi_{j}^-\rangle =m_{j-1,j-1}-m_{j-1,j}+m_{j,j-1}-m_{j,j},\\
\ \ \ \ \ \ \ \ \ \ \ \ \ \ \ \ \ \ \  1\leq i\leq n, \ n+1\leq j\leq 2n.
 \end{array}
 $$}
Because $m_{ij}=0,\text{for} \ i\neq j,$  we can deduce $m_{11}=m_{22}=\cdots=m_{2n+1,2n+1}.$
Hence, $M^{\dagger}M\propto\alpha I $ for some $\alpha\in\mathbb{R}$.

So we can conclude that, the only measurement for Alice that preserves the orthogonality of these states is the trivial measurement.\qed

Now we consider the small set of locally indistinguishable product states in $\mathbb{C}^m\otimes\mathbb{C}^n$. We separate it into three cases  $\mathbb{C}^{2k+1}\otimes\mathbb{C}^{2l+1}, \mathbb{C}^{2k}\otimes\mathbb{C}^{2l+1}$ and $\mathbb{C}^{2k}\otimes\mathbb{C}^{2l}.$

Firstly, in $\mathbb{C}^{2k+1}\otimes\mathbb{C}^{2l+1}$, we construct $6(k+l)-3$ states.
$$
\begin{array}{l}
|\psi_{i}^{\pm}\rangle=|1\rangle(|i\rangle\pm|i+1\rangle),\ \ i=1,3,\ldots,2l-1;\\
|\psi_{i}^{\pm}\rangle=|2k+1\rangle(|i\rangle\pm|i+1\rangle),\ \ i=2,4,\ldots,2l-2;\\
|\psi_{i}\rangle=|k+1\rangle|i\rangle,\ \ i=2,3,\ldots,2l;\\
|\phi_{j}^{\pm}\rangle=(|j\rangle\pm|j+1\rangle)|2l+1\rangle,\ \ j=1,3,\ldots,2k-1;\\
|\phi_{j}^{\pm}\rangle=(|j\rangle\pm|j+1\rangle)|1\rangle,\ \ j=2,4,\ldots,2k-2;\\
|\phi_{j}\rangle=|j\rangle|l+1\rangle,\ \ j=2,3,\ldots,2k.
\end{array}
$$
In {$\mathbb{C}^{5}\otimes\mathbb{C}^{5},$}  these states are plotted as in \textbf{Fig.3:}
$$
\begin{tikzpicture}[
  node distance = 3pt,
  domino/.style = {rectangle, rounded corners = 2mm, draw = black!95, fill = black!20},
  dominoo/.style = {rectangle, rounded corners = 2mm, draw = black!95, fill = black!100}]
  \def\step{20pt};   
  \def\dshort{16pt}; 
  \def\dlong{36pt};  
  \draw[step = \step, red, thin] (0,0) grid (5*\step, 5*\step);
   \foreach \i/\j in {0.5/1,0.5/3,4.5/2,4.5/4}
    \node[dominoo, double, minimum height = \dlong,  minimum width = \dshort] at (\i*\step, \j*\step) {};
   \node[domino, minimum height = \dshort,double, minimum width = \dshort] at (1.5*\step, 2.5*\step) {};
   \node[domino, minimum height = \dshort,double, minimum width = \dshort] at (2.5*\step, 2.5*\step) {};
   \node[domino, minimum height = \dshort,double, minimum width = \dshort] at (3.5*\step, 2.5*\step) {};
   \node[domino, minimum height = \dshort,double, minimum width = \dshort] at (2.5*\step, 1.5*\step) {};
   \node[domino, minimum height = \dshort,double, minimum width = \dshort] at (2.5*\step, 3.5*\step) {};
   \foreach \i/\j/\h in {1/4.5/3,3/4.5/3,2/0.5/3,4/0.5/3}
    \node[dominoo, double , minimum height = \dshort, minimum width = \dlong]  at (\i*\step, \j*\step) {};
   \foreach \i in {1,2,3,4,5}{
    \node (a\i) at (\i*\step-0.5*\step, 3.8)  {$|\i\rangle$};
    \node (b\i) at (-0.35,-\i*\step+5.5*\step) {$|\i\rangle$};
  };
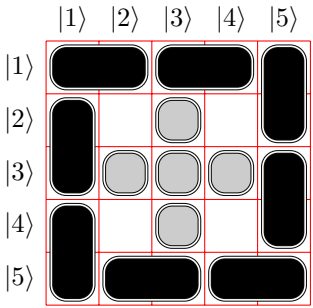
  \node [below = of {1.8,0}] {{\bf Fig.3} \text{Product states representation in } $\mathbb{C}^{5}\bigotimes\mathbb{C}^{5}$};
\end{tikzpicture}
$$
Secondly, in $\mathbb{C}^{2k}\otimes\mathbb{C}^{2l+1}$, we construct $6(k+l)-6$ states:
$$
\begin{array}{l}
|\psi_{i}^{\pm}\rangle=|1\rangle(|i\rangle\pm|i+1\rangle),\ \ i=1,3,\ldots,2l-1;\\
|\psi_{i}^{\pm}\rangle=|2k\rangle(|i\rangle\pm|i+1\rangle),\ \ i=2,4,\ldots,2l-2;\\
|\psi_{i}\rangle=|i\rangle|3\rangle,\ \ i=2,3,\ldots,2k-1;\\
|\phi_{j}^{\pm}\rangle=(|j\rangle\pm|j+1\rangle)|2l+1\rangle,\ \ j=1,3,\ldots,2k-3;\\
|\phi_{j}^{\pm}\rangle=(|j\rangle\pm|j+1\rangle)|1\rangle,\ \ j=2,4,\ldots,2k-4;\\
|\phi_{2k-2}^{\pm}\rangle=(|2k-2\rangle\pm|2k-1\rangle)|2\rangle;\\
|\phi_{2k-1}^{\pm}\rangle=(|2k-1\rangle\pm|2k\rangle)|1\rangle;\\
|\phi_{j}\rangle=|2\rangle|j\rangle,\ \ j=2,3,\ldots,2l.
\end{array}
$$
In {$\mathbb{C}^{6}\otimes\mathbb{C}^{5},$}  these states are plotted as in \textbf{Fig.4:}

$$
\begin{tikzpicture}[
  node distance = 3pt,
  domino/.style = {rectangle, rounded corners = 2mm, draw = black!95, fill = black!20},
  dominoo/.style = {rectangle, rounded corners = 2mm, draw = black!95, fill = black!100}]
  \def\step{20pt};   
  \def\dshort{16pt}; 
  \def\dlong{36pt};  
  \draw[step = \step, red, thin] (0,0) grid (5*\step, 6*\step);
   \foreach \i/\j in {0.5/4,0.5/1,4.5/3,4.5/5,1.5/2}
    \node[dominoo, double, minimum height = \dlong,  minimum width = \dshort] at (\i*\step, \j*\step) {};
   \node[domino, minimum height = \dshort,double, minimum width = \dshort] at (2.5*\step, 1.5*\step) {};
   \node[domino, minimum height = \dshort,double, minimum width = \dshort] at (2.5*\step, 2.5*\step) {};
   \node[domino, minimum height = \dshort,double, minimum width = \dshort] at (2.5*\step, 3.5*\step) {};
   \node[domino, minimum height = \dshort,double, minimum width = \dshort] at (2.5*\step, 4.5*\step) {};
   \node[domino, minimum height = \dshort,double, minimum width = \dshort] at (1.5*\step, 4.5*\step) {};
   \node[domino, minimum height = \dshort,double, minimum width = \dshort] at (3.5*\step, 4.5*\step) {};
   \foreach \i/\j/\h in {1/5.5/3,3/5.5/3,2/0.5/3,4/0.5/3}
    \node[dominoo, double , minimum height = \dshort, minimum width = \dlong]  at (\i*\step, \j*\step) {};
   \foreach \i in {1,2,3,4,5}{
    \node (a\i) at (\i*\step-0.5*\step, 4.6)  {$|\i\rangle$};
  };
   \foreach \i in {1,2,3,4,5,6}{
    \node (b\i) at (-0.35,-\i*\step+6.5*\step) {$|\i\rangle$};
  };
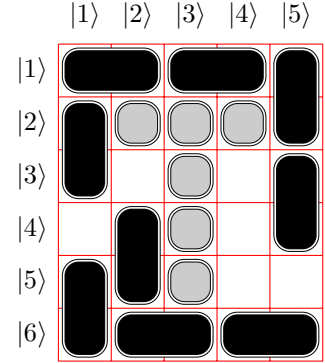
  \node [below = of {1.7,0}] {{\bf Fig.4} \text{Product states representation in } $\mathbb{C}^{6}\bigotimes\mathbb{C}^{4}$};
\end{tikzpicture}
$$
Thirdly, in {$\mathbb{C}^{2k}\otimes\mathbb{C}^{2l}$}, we construct $6(k+l)-9$ states:
$$
\begin{array}{l}
|\psi_{i}^{\pm}\rangle=|1\rangle(|i\rangle\pm|i+1\rangle),\ \ i=1,3,\ldots,2l-3;\\
|\psi_{i}^{\pm}\rangle=|2k\rangle(|i\rangle\pm|i+1\rangle),\ \ i=2,4,\ldots,2l-4;\\
|\psi_{j}\rangle=|j\rangle|3\rangle,\ \ j=3,4,\ldots,2k-1;\\
|\psi_{2l-2}^\pm\rangle=|2k-1\rangle(|2l-2\rangle\pm|2l-1\rangle);\\
|\psi_{2l-1}^{\pm}\rangle=|2k\rangle(|2l-1\rangle\pm|2l\rangle);\\
|\phi_{i}\rangle=|2\rangle|i\rangle,\ \ i=2,3,\ldots,2l-1;\\
|\phi_{j}^{\pm}\rangle=(|j\rangle\pm|j+1\rangle)|2l\rangle,\ \ j=1,3,\ldots,2k-3;\\
|\phi_{j}^{\pm}\rangle=(|j\rangle\pm|j+1\rangle)|1\rangle,\ \ j=2,4,\ldots,2k-4;\\
|\phi_{2k-2}^{\pm}\rangle=(|2k-2\rangle\pm|2k-1\rangle)|2\rangle;\\
|\phi_{2k-1}^{\pm}\rangle=(|2k-1\rangle\pm|2k\rangle)|1\rangle.\\
\end{array}
$$
In {$\mathbb{C}^{6}\otimes\mathbb{C}^{6},$}  these states are plotted as in \textbf{Fig.5:}
$$
\begin{tikzpicture}[
  node distance = 3pt,
  domino/.style = {rectangle, rounded corners = 2mm, draw = black!95, fill = black!20},
  dominoo/.style = {rectangle, rounded corners = 2mm, draw = black!95, fill = black!100}]
  \def\step{20pt};   
  \def\dshort{16pt}; 
  \def\dlong{36pt};  
  \draw[step = \step, red, thin] (0,0) grid (6*\step, 6*\step);
   \foreach \i/\j in {0.5/1,0.5/4,1.5/2,5.5/3,5.5/5}
    \node[dominoo, double, minimum height = \dlong,  minimum width = \dshort] at (\i*\step, \j*\step) {};
   \node[domino, minimum height = \dshort,double, minimum width = \dshort] at (1.5*\step, 4.5*\step) {};
   \node[domino, minimum height = \dshort,double, minimum width = \dshort] at (2.5*\step, 4.5*\step) {};
   \node[domino, minimum height = \dshort,double, minimum width = \dshort] at (3.5*\step, 4.5*\step) {};
   \node[domino, minimum height = \dshort,double, minimum width = \dshort] at (4.5*\step, 4.5*\step) {};
   \node[domino, minimum height = \dshort,double, minimum width = \dshort] at (2.5*\step, 3.5*\step) {};
   \node[domino, minimum height = \dshort,double, minimum width = \dshort] at (2.5*\step, 2.5*\step) {};
   \node[domino, minimum height = \dshort,double, minimum width = \dshort] at (2.5*\step, 1.5*\step) {};
   \foreach \i/\j/\h in {1/5.5/3,3/5.5/3,4/1.5/3,2/0.5/3,5/0.5/3}
    \node[dominoo, double , minimum height = \dshort, minimum width = \dlong]  at (\i*\step, \j*\step) {};
   \foreach \i in {1,2,3,4,5,6}{
    \node (a\i) at (\i*\step-0.5*\step, 4.6)  {$|\i\rangle$};
    \node (b\i) at (-0.35,-\i*\step+6.5*\step) {$|\i\rangle$};
  };
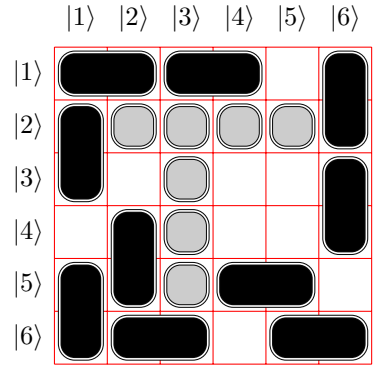
 \node [below = of {1.9,0}] {{\bf Fig.5} \text{Product states representation in } $\mathbb{C}^{6}\bigotimes\mathbb{C}^{6}$};
\end{tikzpicture}
$$

\emph{Theorem 2.}
 In $\mathbb{C}^m\otimes\mathbb{C}^n$, there are $3(m+n)-9$ product  states that are  LOCC indistinguishable and these states
are constructed as above, and we denote the set as $T$.

\emph{Proof:} Now we just give the proof of the second  case, others are just as similar. If these states are LOCC distinguishable when Alice goes first, then Alice can perform a nontrivial measurement $\{M_i^A\}$ which preserves the orthogonality of these states. Next we prove that for any matrix $M$ that preserves the orthogonality of the given states, then $M^{\dagger}M\propto \alpha I$ for some $\alpha\in \mathbb{R}.$
We can suppose  $M^{\dagger}M=(m_{ij}).$  Because the label $i$ of $|\psi_{i}\rangle$ runs from $2$,
 to $2k-1$, we can define $|\psi_{1}\rangle=|\psi_{1}^+\rangle,|\psi_{2k}\rangle=|\psi_{2k}^+\rangle$. Then by  the orthogonality of $M\otimes I|\psi_{i}\rangle$ and $M\otimes I|\psi_{j}\rangle$, we have
$$0=\langle\psi_{j}|M^{\dagger}M\otimes I|\psi_{i}\rangle=m_{ji}, i\neq j.$$
Then we have $m_{ij}=0  \text{ for } i\neq j.$ So the matrix $M$ is a diagonal matrix. Next we show all the diagonal elements in the matrix $M$ are equal to each other.
By the orthogonality of $M\otimes I|\phi_{j}^{+}\rangle$ and $M\otimes I|\phi_{j}^{-}\rangle$ $(j=1,2,\ldots,2k-1)$, we have
$$0=\langle\phi_{j}^{+}|M^{\dagger}M\otimes I|\phi_{j}^-\rangle=m_{j,j}-m_{j,j+1}+m_{j+1,j}-m_{j+1,j+1}.$$
Then we have $m_{j,j}=m_{j+1,j+1}$ for $m_{j+1,j}=m_{j,j+1}=0$ where $j=1,2,\ldots,2k-1.$
So we deduce $m_{11}=m_{22}=\cdots=m_{2k,2k}.$ Hence  $M^{\dagger}M\propto\alpha I $ for some $\alpha\in\mathbb{R}.$

So we can deduce that  the only measurement for Alice that preserves the orthogonality of these states is the trivial measurement.
Similarly, if Bob goes  first, we can also prove that the measurement Bob can do is also a trivial measurement.  \qed

We can easily notice that the set $T$ of product states we constructed in theorem 1 and theorem 2 is not an UPB.

\emph{Lemma.}  The set $T$ of product states we constructed in theorem 1 and theorem 2 is not an UPB. Moreover, the set $T $ can be extended to an orthogonal product base.

\emph{Proof }: Because the states in theorem 1 are only a subset of a product base constructed in \cite{Zhang14}, so it can be extended to a product base obviously. Next, we consider the rectangle representation of the product states we constructed in theorem 2. For the case in $\mathbb{C}^m\otimes\mathbb{C}^n$, there are $mn-3(m+n)+9$ white squares. Each square represents a product state. Then we have $mn-3(m+n)+9$  product states, and we denote this set as $S$. It is not difficult to show that the states in $S\cup T$ are mutually orthogonal. And there are exactly $mn$ elements in the set $S\cup T$ . So we can conclude that $S\cup T$ is an orthogonal product base in $\mathbb{C}^m\otimes\mathbb{C}^n$. \qed

\emph{Theorem 3.} The LOCC indistinguishable product states we constructed in theorem 1 and theorem 2 can
be distinguished by separable measurements.

\emph{Proof:} In $\mathbb{C}^m\otimes\mathbb{C}^n$, we have construct $N=3(m+n)+9$ product states. We can denote it as $|\psi_1\rangle,
|\psi_2\rangle, ..., |\psi_N\rangle$. From the above lemma, these $N$ product states can be extended to an orthogonal product base,
so we can denote the other $mn-N$ product states as $|\psi_{N+1}\rangle, |\psi_{N+2}\rangle, ..., |\psi_{mn}\rangle.$ Now we define a measurement
$\{M_i\}_{i=1}^{mn} $, where $M_i=|\psi_i\rangle \langle \psi_i|.$  Because the set $\{|\psi_i\}_{i=1}^{mn}$ is an orthogonal normal base of $\mathbb{C}^m\otimes\mathbb{C}^n$, the completeness equation holds. That is,
                $$\sum_{i=1}^{mn}M_i=\sum_{i=1}^{mn}|\psi_i\rangle \langle \psi_i|=I.$$
As any $|\psi_i\rangle$ is a product state, we can conclude that $M_i$ is separable. So $\{M_i\}_{i=1}^{mn} $ is a separable measurement.

Now we state that these $N$ product states  $|\psi_1\rangle,$ $|\psi_2\rangle, ..., |\psi_N\rangle$ can be distinguished by the above separable measurement. First, we have the following equations:
{\small
$$\langle \psi_j |M_i|\psi_j\rangle = \langle \psi_j |\psi_i\rangle \langle \psi_i|\psi_j\rangle = \delta_{ij}, 1 \leq i \leq mn, 1 \leq j \leq N.$$
}
So the outcome could not be greater than $N$. Moreover, if the outcome is $i$, then we can correctly conclude that the given state is $|\psi_i\rangle$. Hence the set of product states $|\psi_1\rangle,|\psi_2\rangle, ..., |\psi_N\rangle$  can be distinguished by separable measurement. \qed

\section{conclusion}

We study the locally indistinguishable problem of orthogonal pure product states in bipartite quantum system. First, we find a small set of pure product states
which is LOCC indistinguishable in $\mathbb{C}^d\otimes\mathbb{C}^d$. This gives an answer of the question raised by the authors
in Ref.\cite{Zhang14}. Then we generalize our method to  arbitrary bipartite quantum system $\mathbb{C}^m\otimes\mathbb{C}^n$ and  construct a set contains only $3(m+n)-9$ pure product states which are LOCC indistinguishable. What surprising us is that the set of $3(m+n)-9$ pure product states can
be distinguished by separable measurements for it can be extended to a orthogonal product base. This shows that separable operations are stronger than the local operations and classical communication. We hope that our results will help us a better understanding the ¡±nonlocality without entanglement¡±.

\bigskip

\noindent \textbf{Acknowledgments }  All the authors thanks for the referee's suggestions to improve the quality of this paper.
This work is supported by the NSFC through Grants
No. 11475178 and No. 11275131.

\end{document}